\documentclass{aastex7}
\usepackage{graphicx} 
\usepackage{amsmath,amssymb} 
\usepackage{longtable}

\begin{document}

\author[0000-0003-2858-9657]{John L. Tonry}
\affiliation{Institute for Astronomy, University of Hawaii, 2680 Woodlawn Drive, Honolulu, HI 96822, USA}
\email[show]{tonry@hawaii.edu}

\author[0000-0002-7034-148X]{Larry Denneau, Jr}
\affiliation{Institute for Astronomy, University of Hawaii, 2680 Woodlawn Drive, Honolulu, HI 96822, USA}
\email{denneau@hawaii.edu}

\author[orcid=0000-0002-8134-2592]{Miguel R. Alarc\'on}
\affiliation{Instituto de Astrof\'{\i}sica de Canarias (IAC), C/V\'{\i}a L\'actea s/n, 38205 La Laguna, Tenerife, Spain}
\affiliation{Departamento de Astrof\'{\i}sica, Universidad de La Laguna (ULL), 38206 La Laguna, Tenerife, Spain}
\email{mra@iac.es}

\author[0000-0003-3068-4258]{Alejandro Clocchiatti}
\affiliation{Pontificia Universidad Catolica de Chile and Millennium Institute of Astrophysics}
\email{aclocchi@uc.cl}

\author[0000-0002-9986-3898]{Nicolas Erasmus}
\affiliation{South African Astronomical Observatory, Cape Town, 7925, South Africa and Department of Physics, Stellenbosch University, Stellenbosch, 7602, South Africa}
\email{nerasmus@saao.ac.za}

\author[0000-0003-0250-9911]{Alan Fitzsimmons}
\affiliation{Astrophysics Research Centre, Queen's University Belfast, Belfast BT7 1NN, UK}
\email{a.fitzsimmons@qub.ac.uk}

\author[0000-0001-5346-436X]{Javier Licandro}
\affiliation{Instituto de Astrof\'{\i}sica de Canarias (IAC), C/V\'{\i}a L\'actea s/n, 38205 La Laguna, Tenerife, Spain}
\affiliation{Departamento de Astrof\'{\i}sica, Universidad de La Laguna (ULL), 38206 La Laguna, Tenerife, Spain}
\email{jlicandr@iac.es}

\author[0000-0002-2058-5670]{Karen J. Meech}
\affiliation{Institute for Astronomy, University of Hawaii, 
2680 Woodlawn Drive, Honolulu, HI 96822, USA}
\email{meech@hawaii.edu}  

\author[0000-0001-5016-3359]{Robert Siverd}
\affiliation{Institute for Astronomy, University of Hawaii, 2680 Woodlawn Drive, Honolulu, HI 96822, USA}
\email{hweiland@hawaii.edu}

\author[0000-0003-1847-9008]{Henry Weiland}
\affiliation{Institute for Astronomy, University of Hawaii, 2680 Woodlawn Drive, Honolulu, HI 96822, USA}
\email{hweiland@hawaii.edu}

\title{ATLAS Photometry of Interstellar Object 3I/ATLAS}

\begin{abstract}
We present calibrated ATLAS photometry of the interstellar comet 3I/ATLAS (C/2025 N1) from 28 March through 29 Aug 2025, obtained with the five-site, robotic ATLAS network in the $c$ (420-650~nm), $o$ (560-820~nm), and Teide $w$ (420-720~nm) bands. Stacked difference images yield reliable light curves measured in four fixed apertures that capture the evolving coma. We observe 3I/ATLAS transitioning in color from red $(c-o)\approx0.7$ before MJD 60860 to near-solar $(c-o)\approx0.3$ after MJD 60870, coincident with the appearance of a prominent anti-solar tail. The absolute magnitude curve $H(t)$ shows a slope break near MJD 60890 at $r\sim3.3$~au from $-0.035$ to $-0.012$~mag/day, or in terms of coma cross section as a function of heliocentric distance, $r^{-3.9}$ to $r^{-1.1}$.  We release the aperture photometry with geometry and uncertainties to enable cross-instrument synthesis of 3I/ATLAS activity and color evolution.
\end{abstract}


\section{Introduction}

The third example of an object from another star passing through the solar system was discovered by the ATLAS system on 1 July 2025.  Immediately recognized as following a hyperbolic orbit, it was named C/2025~N1 (ATLAS) or 3I/ATLAS and became the target of intensive observing campaigns.  Additional detections of 3I/ATLAS prior to 1 July 2025 were also found in various images and contributed to the understanding of how 3I/ATLAS grew a coma as it approached the Sun.  \cite{Seligman2025} summarizes the state of knowledge of 3I/ATLAS in those early days in July.

Two other interstellar objects have been identified  before the discovery of 3I/ATLAS, 1I/Oumuamua \citep{Williams17} and 2I/Borisov \citep{borisov_2I_cbet}.  The stark difference between 1I/Oumuamua ($\sim$100~m, no volatiles) compared with 2I/Borisov and 3I/ATLAS (few-km, copious cometary activity) is extremely interesting but as yet poorly understood.  The prospects for finding more examples is very good.  The ATLAS survey would have discovered 2I/Borosov a few weeks after the actual discovery, and ATLAS has such a simple survey pattern (all-sky more than 50$^\circ$ from the sun, 4 times per night, limiting magnitude $\sim$19.3, weather and moon permitting) it is easy to predict the rate that ATLAS should discover interstellar objects for a given density.  

We presented a calculation \citep{Seligman2025} that estimates the ATLAS discovery cross section multiplied by typical interstellar object velocity as $600$~au$^3$/yr for objects of the $\sim$10~km effective size of 3I/ATLAS including its coma, so the detection of two such objects in 6 years by ATLAS implies a local density of $\sim$$3\times10^{-4}$~au$^{-3}$.  The volume inside of Jupiter probably contains a new $\sim$10~km (effective) scale interstellar object every 5 years.

An important question is why we have not yet seen interstellar objects intermediate in size between 1I/Oumuamua and 2I/Borisov and 3I/ATLAS.  The ATLAS detection cross section (and therefore the detection rate for a constant density) is approximately inversely proportional to object diameter down to $\sim$1~km effective size, and populations of asteroids and comets in the solar system grow faster than that with diminishing size, so if the population of interstellar objects has a similar distribution to solar system objects we should have already discovered quite a few interstellar objects of effective size $\sim$1~km.

The subject of this paper is more mundane than these population questions, however.  Since ATLAS observes all the sky all the time we have likely detections of 3I/ATLAS as early as April 2025, and continuously ever since.  During this span of time the 3I/ATLAS coma has changed dramatically from a red ``rubble fountain'' pointing towards the sun to a fully developed, white comet tail pointing away.  The ATLAS instruments do not have the resolution to isolate the comet nucleus (even HST and JWST cannot do that), so we present the ATLAS photometry through 30 Aug 2025 in a set of 4 concentric apertures that capture more and more coma brightness.

\section{Observations and Data}

The ATLAS system \citep{Tonry2018a} operates at 5 sites, Haleakala (site ``02'', MPC T08) and Mauna Loa (site ``01'', MPC T05) in Hawaii, Sutherland Station (site ``03'', MPC M22) in South Africa run by the South Africa Astronomical Observatory, Rio Hurtado (site ``04'', MPC W68) in Chile (where the discovery was made) run by Obstech, and Tenerife (site ``05'', MPC R17) \citep{2023sndd.confE...2L}.  These facilities operate robotically without human intervention, and the images pass through an automatic reduction pipeline that includes astrometric and photometric calibration, differencing against an all-sky static ``wallpaper'' from many previous images, detection by 5 different algorithms, detection classification, linkage of detections into moving objects, static variables, and static transients, winnowing by a convolutional neural net, presentation to humans for final vetting, and submission to either the Minor Planet Center or the Transient Name Server.  Nothing is held back, all data are immediately available to the world via these object servers and the ATLAS forced photometry server\footnote{\tt https://fallingstar-data.com/forcedphot/} \citep{forced2021}.

\begin{figure}[h]
\begin{center}$
\begin{array}{cc}
\includegraphics[width=0.4\linewidth]{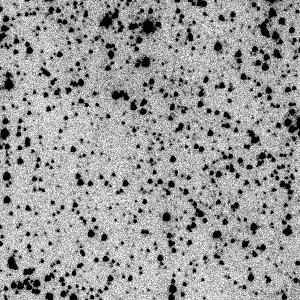}
\qquad
\includegraphics[width=0.4\linewidth]{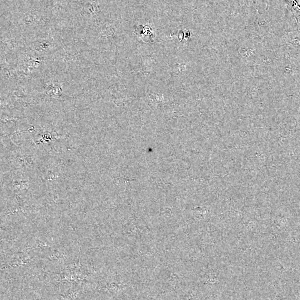}
\end{array}$
\end{center}
\caption{The discovery $o$ band image of 3I/ATLAS from MJD 60957 is shown on the left centered on 3I/ATLAS, and the four difference images from that night are stacked on the right.  The 9.3\arcmin\ image on the left has north up, east left with the native 1.86\arcsec pixels; the 10\arcmin\ image on the right has 2\arcsec\ rebinned pixels; black indicates surface brightness brighter than 21~AB~arcsec$^{-2}$.}
\label{fig:disc}
\end{figure}

\begin{figure}[h]
\begin{center}$
\begin{array}{ccdd}
\includegraphics[width=0.2\linewidth]{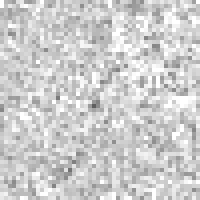}
\quad
\includegraphics[width=0.2\linewidth]{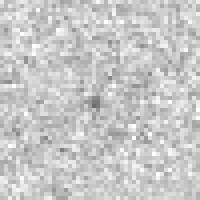}
\quad
\includegraphics[width=0.2\linewidth]{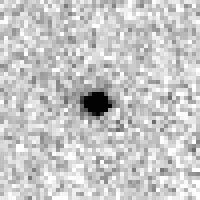}
\quad
\includegraphics[width=0.2\linewidth]{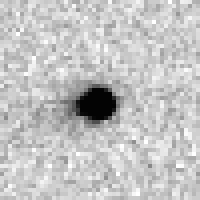}
\end{array}$
\end{center}
\caption{These images illustrate the evolution of 3I/ATLAS. From left to right are the $c$ band stack from MJD 60818–60826 (28 May)
the $o$ band stack from 60817–60828 (26 May), the epoch at 60884 (27 Jul) just before the $H$ band slope change described below, and the epoch at 60905 (17 Aug) when the eastward, anti-solar tail becomes prominent. The fields are 100\arcsec\ with 2\arcsec\ pixels, north up and east left, black indicates surface brightness brighter than 23.2 AB arcsec$^{-2}$.}
\label{fig:evolve}
\end{figure}

We normally collect 4 images each night of each field because linkage of moving objects is {\it much} simpler if 4 detections exist within a short period of time, although there are certainly methods to link over longer time intervals \citep{Tonry2023, Heinze2022, Holman2018}.  Images of objects that vary more slowly than the $\sim$40~min spent collecting the data can be rebinned onto a common coordinate system (moving for a moving object) and stacked for improved signal to noise.  The difference imaging stage is particularly important in crowded fields such as the galactic plane where 3I/ATLAS was discovered, but it requires great care in astrometric solutions, $\sim$0.02 pixel or better.  Similarly, the
stacking of the earliest images from ATLAS had to wait for the extrapolation of the 3I/ATLAS orbit to attain sub-arcsecond accuracy. 
Figures~\ref{fig:disc} and \ref{fig:evolve} show the appearance of 3I/ATLAS in the stacked images.  

Photometry is reported in three filters: ATLAS $c$ band (420--650~nm), ATLAS $o$ band (560--820~nm), and the Teide $w$ band (approximately 420--720~nm).  \cite{Tonry2018a} describes the $c$ and $o$ bandpasses in detail, but the $w$ bandpass has not been thoroughly characterized.  The color conversion between $c$, $o$, and $w$ to the Pan-STARRS $g$, $r$, $i$, or $V$ bandpass \citep{Tonry2012} depends on the spectral energy distribution, but integrating $c$, $o$, $g$, $r$, $i$, and $V$ against a wide range of stellar SEDs yields tight relationships
\begin{align}
(r-o) &= -0.022 + 0.282\, (c-o) \\
(V-o) &= -0.007 + 0.884\, (c-o) \\
(g-i) &= -0.030 + 1.915\, (c-o),
\label{eq:color}
\end{align}
over the range $0<(c-o)<0.8$.  It is probably adequate to take $w\approx V$ (bearing in mind that $w$ extends well to the blue and red of $V$ band and is on the AB system).  Alternatively we find empirically that $(w-g)\approx-0.6\,(g-r)$.  As discussed below, we appear to have seen a significant color change in 3I/ATLAS from $(c-o)\approx0.7$ to $(c-o)\approx0.3$, so due consideration must be taken for the time varying color of the SED if converting from observed bandpasses.  All photometry is tied to Refcat2 \citep{Tonry2018b} which is founded on Pan-STARRS and Gaia.

Unfortunately most of the photometry reported in the literature does not provide enough information for intercomparison.  In this paper we use $c$ and $o$ bandpasses whose detailed profiles are published, we specify the reference to which our photometry zeropoints are tied, the apertures for the photometry are exactly defined (although seeing obviously plays a role in smaller apertures), we illustrate how our difference images credibly remove stellar contamination, and extremely importantly, our background level is derived from a constant plus $\rho^{-3}$ fit to median ring surface brightness between radii 64--128\arcsec.

The angular speed of 3I/ATLAS during this time was about 1.4\arcsec/min and all exposure times except for 03a60905 and 04a60914 were 30~sec, so the blurring from motion is about 1/3 of a 2\arcsec\ pixel and not significant.

Table~\ref{tab:phot} gives the magnitude measured for 3I/ATLAS in four apertures drawn from stacked difference images with 2\arcsec\ pixels.  This flux conserving stacking was aligned on the orbit of 3I/ATLAS, without any reference to flux in the images, so the center of the central pixel corresponds to the location of 3I/ATLAS at the midpoint of the exposure.
``StackID'' codes site, instrument, and integer MJD night number, where a hyphen indicates the MJD range if more than $\sim$1~hour, ``MJD'' is the actual mean time of the exposure (1.0 July 2025 is MJD 60857.0), ``N'' is the number of exposures contributing to the stack, 
``m2'' is the central 2\arcsec\ pixel,
``m6'' is the 9 pixel 6$\times$6\arcsec\ square aperture,
``m10'' is the 21 pixels of the 5$\times$5 pixel square less each corner pixel to make a rounded aperture of diameter $\sim$10\arcsec, ``m14'' is the 37 pixels of the 7$\times$7 pixel square removing three pixels from each corner making a rounded aperture of diameter $\sim$14\arcsec,
``dm6'' is the uncertainty of the ``m6'' magnitude (other uncertainties can be scaled by square root of area), ``r'' is the heliocentric distance [au], ``Delta'' is the Earth distance [au], ``sto'' is the Sun-Target-Observer angle [deg],
and ``HM'' is the ``Halley-Marcus'' phase function for comets evaluated at sto, assembled by D. Schleicher\footnote{\tt https://asteroid.lowell.edu/comet/dustphase}.  The asterisk on 05r60868 and 05r60883 indicates potentially unreliable photometry, as described below.
The typical point spread function for these observations is about 4\arcsec\ full width half maximum (the $o$ and $c$ images have 1.86\arcsec/pixel and the $w$ band 1.25\arcsec/pixel) so the central pixel is significantly diminished by the seeing relative to infinite resolution.  These data are available in digital form at {\tt http://astroportal.ifa.hawaii.edu/atlas/3I-photometry}.

\begin{longtable}{lccrccccccccc}
\caption{ATLAS Photometry of 3I/ATLAS.} \\
\hline
StackID & MJD & filt & N & m2 & m6 & m10 & m14 & dm6 & r & Delta & sto & HM \\
\hline
\endhead
\hline
\endfoot
60760-60764o & 60762.4563 & $o$ & 12 & 24.14 & 23.74 & 21.71 & 21.06 & 1.88 & 7.705 & 7.840 &  7.29 &  0.30 \\
60762-60766c & 60764.7003 & $c$ & 12 & 23.39 & 22.49 & 22.78 & 22.78 & 0.57 & 7.628 & 7.725 &  7.42 &  0.31 \\
60791-60792c & 60791.5831 & $c$ &  8 & 23.27 & 22.16 & 21.76 & 21.94 & 0.43 & 6.718 & 6.340 &  8.20 &  0.34 \\
60786-60798o & 60792.5380 & $o$ & 29 & 23.70 & 21.61 & 21.96 & 21.86 & 0.19 & 6.685 & 6.292 &  8.17 &  0.34 \\
60817-60828o & 60821.2754 & $o$ & 17 & 22.73 & 20.63 & 20.14 & 19.90 & 0.07 & 5.714 & 4.897 &  6.43 &  0.27 \\
60818-60826c & 60823.0204 & $c$ & 17 & 22.83 & 21.29 & 20.86 & 21.25 & 0.17 & 5.655 & 4.818 &  6.22 &  0.26 \\
04a60843     & 60843.2952 & $o$ &  4 & 20.98 & 19.43 & 18.92 & 18.81 & 0.08 & 4.973 & 3.976 &  2.55 &  0.11 \\
05r60845     & 60845.0836 & $w$ &  4 & 20.86 & 19.31 & 18.81 & 18.66 & 0.05 & 4.913 & 3.910 &  2.14 &  0.09 \\
05r60848     & 60848.0824 & $w$ &  4 & 20.88 & 19.13 & 18.86 & 18.76 & 0.04 & 4.812 & 3.802 &  1.47 &  0.07 \\
04a60850     & 60850.4007 & $o$ &  4 & 20.47 & 18.62 & 18.23 & 18.09 & 0.04 & 4.735 & 3.722 &  1.09 &  0.05 \\
05r60851     & 60851.0416 & $w$ &  3 & 21.12 & 19.18 & 18.70 & 18.52 & 0.04 & 4.714 & 3.700 &  1.03 &  0.05 \\
03a60851     & 60851.0812 & $c$ &  5 & 21.72 & 19.65 & 19.06 & 18.73 & 0.06 & 4.712 & 3.698 &  1.03 &  0.05 \\
04a60853     & 60853.2509 & $o$ &  4 & 20.26 & 18.31 & 17.86 & 17.80 & 0.05 & 4.640 & 3.626 &  1.10 &  0.05 \\
02a60854     & 60854.4585 & $c$ &  4 & 21.11 & 19.12 & 18.61 & 18.47 & 0.05 & 4.599 & 3.587 &  1.30 &  0.06 \\
05r60855     & 60855.0381 & $w$ &  4 & 20.45 & 18.57 & 18.17 & 17.98 & 0.03 & 4.580 & 3.569 &  1.42 &  0.06 \\
03a60856     & 60855.9503 & $c$ &  4 & 21.02 & 19.03 & 18.52 & 18.31 & 0.02 & 4.550 & 3.540 &  1.64 &  0.07 \\
04a60857     & 60857.2336 & $o$ &  4 & 20.10 & 18.34 & 17.97 & 17.84 & 0.02 & 4.507 & 3.500 &  1.99 &  0.09 \\
05r60858     & 60858.0347 & $w$ &  4 & 20.28 & 18.50 & 18.23 & 18.16 & 0.02 & 4.480 & 3.475 &  2.23 &  0.10 \\
04a60858     & 60858.2348 & $o$ &  4 & 20.45 & 18.45 & 17.95 & 17.78 & 0.04 & 4.473 & 3.469 &  2.30 &  0.10 \\
03a60859     & 60858.9859 & $c$ &  4 & 20.72 & 18.81 & 18.35 & 18.20 & 0.02 & 4.449 & 3.447 &  2.53 &  0.11 \\
04a2300      & 60860.9796 & $o$ & 18 & 20.24 & 18.32 & 17.92 & 17.73 & 0.02 & 4.382 & 3.388 &  3.18 &  0.14 \\
05r60861     & 60861.0135 & $w$ &  4 & 20.31 & 18.47 & 18.06 & 17.98 & 0.03 & 4.381 & 3.387 &  3.19 &  0.14 \\
04a0000      & 60861.0213 & $o$ & 21 & 20.15 & 18.27 & 17.83 & 17.69 & 0.02 & 4.381 & 3.387 &  3.19 &  0.14 \\
04a0100      & 60861.0620 & $o$ & 18 & 20.14 & 18.29 & 17.85 & 17.72 & 0.02 & 4.379 & 3.386 &  3.21 &  0.14 \\
04a0200      & 60861.1044 & $o$ & 16 & 20.21 & 18.27 & 17.87 & 17.75 & 0.02 & 4.378 & 3.384 &  3.22 &  0.14 \\
04a0300      & 60861.1450 & $o$ & 17 & 20.25 & 18.30 & 17.87 & 17.72 & 0.02 & 4.377 & 3.383 &  3.23 &  0.14 \\
04a0400      & 60861.1875 & $o$ & 18 & 20.20 & 18.24 & 17.82 & 17.65 & 0.02 & 4.375 & 3.382 &  3.25 &  0.14 \\
04a0500      & 60861.2296 & $o$ & 16 & 20.21 & 18.28 & 17.85 & 17.72 & 0.02 & 4.374 & 3.381 &  3.26 &  0.14 \\
04a0600      & 60861.2717 & $o$ & 14 & 20.26 & 18.34 & 17.87 & 17.69 & 0.02 & 4.372 & 3.380 &  3.28 &  0.14 \\
02a0600      & 60861.2733 & $o$ & 19 & 20.12 & 18.29 & 17.87 & 17.68 & 0.02 & 4.372 & 3.380 &  3.28 &  0.14 \\
02a0700      & 60861.3128 & $o$ & 23 & 20.27 & 18.31 & 17.91 & 17.76 & 0.01 & 4.371 & 3.378 &  3.29 &  0.14 \\
04a0700      & 60861.3128 & $o$ & 14 & 20.25 & 18.37 & 17.93 & 17.80 & 0.02 & 4.371 & 3.378 &  3.29 &  0.14 \\
04a0800      & 60861.3508 & $o$ & 12 & 20.22 & 18.38 & 17.88 & 17.64 & 0.03 & 4.370 & 3.377 &  3.30 &  0.14 \\
02a0800      & 60861.3556 & $o$ & 25 & 20.27 & 18.34 & 17.91 & 17.76 & 0.01 & 4.369 & 3.377 &  3.31 &  0.14 \\
02a0900      & 60861.3953 & $o$ & 24 & 20.30 & 18.36 & 17.94 & 17.79 & 0.01 & 4.368 & 3.376 &  3.32 &  0.14 \\
02a1000      & 60861.4382 & $o$ & 22 & 20.31 & 18.34 & 17.89 & 17.73 & 0.02 & 4.367 & 3.375 &  3.33 &  0.14 \\
02a1100      & 60861.4789 & $o$ & 20 & 20.22 & 18.39 & 17.93 & 17.72 & 0.02 & 4.365 & 3.374 &  3.35 &  0.15 \\
02a1200      & 60861.5204 & $o$ & 17 & 20.41 & 18.40 & 17.91 & 17.73 & 0.02 & 4.364 & 3.373 &  3.36 &  0.15 \\
04a60862     & 60862.1202 & $o$ &  4 & 20.37 & 18.42 & 17.84 & 17.62 & 0.03 & 4.344 & 3.355 &  3.57 &  0.16 \\
05r60868*    & 60868.0385 & $w$ &  4 & 19.77 & 17.99 & 17.68 & 17.57 & 0.04 & 4.148 & 3.199 &  5.72 &  0.24 \\
04a60868     & 60868.1667 & $o$ &  4 & 20.00 & 18.14 & 17.64 & 17.54 & 0.04 & 4.143 & 3.195 &  5.78 &  0.25 \\
03a60869     & 60868.9227 & $o$ &  4 & 20.10 & 18.08 & 17.57 & 17.38 & 0.05 & 4.119 & 3.177 &  6.06 &  0.26 \\
04a60869     & 60869.2095 & $o$ &  4 & 19.95 & 18.08 & 17.60 & 17.43 & 0.03 & 4.109 & 3.170 &  6.17 &  0.26 \\
04a60870     & 60870.1940 & $o$ &  7 & 19.91 & 18.05 & 17.62 & 17.39 & 0.03 & 4.076 & 3.146 &  6.56 &  0.28 \\
05r60871     & 60870.9906 & $w$ &  4 & 20.23 & 18.32 & 17.90 & 17.72 & 0.02 & 4.050 & 3.127 &  6.87 &  0.29 \\
03a60871     & 60871.0092 & $o$ &  7 & 20.08 & 18.19 & 17.78 & 17.63 & 0.03 & 4.049 & 3.127 &  6.88 &  0.29 \\
03a60872     & 60872.0120 & $o$ &  7 & 19.90 & 18.07 & 17.62 & 17.39 & 0.02 & 4.016 & 3.104 &  7.27 &  0.30 \\
03a60873     & 60872.8108 & $o$ &  2 & 19.95 & 18.07 & 17.64 & 17.46 & 0.02 & 3.990 & 3.086 &  7.58 &  0.32 \\
05r60874     & 60873.9749 & $w$ &  4 & 20.28 & 18.34 & 17.85 & 17.63 & 0.02 & 3.951 & 3.060 &  8.06 &  0.33 \\
04a60874     & 60874.1477 & $c$ &  4 & 20.17 & 18.26 & 17.84 & 17.67 & 0.02 & 3.945 & 3.056 &  8.13 &  0.34 \\
03a60875     & 60874.8838 & $o$ &  4 & 19.92 & 17.94 & 17.49 & 17.29 & 0.01 & 3.922 & 3.041 &  8.42 &  0.35 \\
03a60876     & 60875.8850 & $o$ &  5 & 19.98 & 17.89 & 17.35 & 17.15 & 0.02 & 3.888 & 3.020 &  8.83 &  0.36 \\
03a60877     & 60876.7542 & $o$ &  2 & 19.77 & 17.80 & 17.26 & 17.06 & 0.02 & 3.860 & 3.002 &  9.18 &  0.38 \\
05r60877     & 60876.9726 & $w$ &  4 & 19.77 & 17.92 & 17.55 & 17.37 & 0.01 & 3.852 & 2.998 &  9.28 &  0.38 \\
04a60877     & 60877.1447 & $c$ &  4 & 19.98 & 18.11 & 17.65 & 17.44 & 0.03 & 3.847 & 2.995 &  9.34 &  0.38 \\
04a60878     & 60878.1331 & $c$ &  4 & 20.06 & 18.09 & 17.64 & 17.48 & 0.02 & 3.814 & 2.975 &  9.75 &  0.40 \\
03a60880     & 60879.9488 & $o$ &  4 & 19.58 & 17.66 & 17.19 & 17.02 & 0.01 & 3.755 & 2.941 & 10.50 &  0.42 \\
05r60880     & 60879.9681 & $w$ &  4 & 20.30 & 18.22 & 17.68 & 17.44 & 0.01 & 3.754 & 2.941 & 10.51 &  0.42 \\
04a60881     & 60881.1325 & $c$ &  4 & 19.82 & 17.93 & 17.50 & 17.32 & 0.01 & 3.716 & 2.920 & 10.99 &  0.44 \\
03a60883     & 60882.8439 & $o$ &  4 & 19.46 & 17.53 & 17.06 & 16.88 & 0.01 & 3.660 & 2.890 & 11.70 &  0.47 \\
05r60883*    & 60882.9651 & $w$ &  8 & 20.06 & 18.06 & 17.52 & 17.28 & 0.01 & 3.656 & 2.888 & 11.75 &  0.47 \\
03a60884     & 60883.8467 & $o$ &  4 & 19.43 & 17.47 & 17.01 & 16.83 & 0.01 & 3.627 & 2.874 & 12.11 &  0.48 \\
04a60885     & 60885.0847 & $c$ &  4 & 19.71 & 17.84 & 17.42 & 17.21 & 0.01 & 3.587 & 2.854 & 12.62 &  0.50 \\
05r60886     & 60885.9606 & $w$ &  4 & 19.44 & 17.63 & 17.24 & 17.10 & 0.01 & 3.558 & 2.841 & 12.98 &  0.51 \\
02a60886     & 60886.3258 & $o$ &  4 & 19.29 & 17.33 & 16.85 & 16.67 & 0.01 & 3.546 & 2.835 & 13.13 &  0.51 \\
05r60889     & 60888.9049 & $w$ &  4 & 19.66 & 17.72 & 17.19 & 16.98 & 0.01 & 3.462 & 2.799 & 14.17 &  0.55 \\
04a60894     & 60894.1118 & $o$ &  4 & 19.00 & 17.08 & 16.58 & 16.33 & 0.03 & 3.294 & 2.735 & 16.21 &  0.61 \\
03a60895     & 60894.8868 & $o$ &  4 & 19.02 & 17.07 & 16.49 & 16.35 & 0.03 & 3.269 & 2.727 & 16.50 &  0.62 \\
05r60896     & 60895.9044 & $w$ &  4 & 19.19 & 17.29 & 16.76 & 16.61 & 0.03 & 3.236 & 2.716 & 16.88 &  0.63 \\
05r60900     & 60899.8984 & $w$ &  4 & 19.14 & 17.20 & 16.74 & 16.57 & 0.01 & 3.109 & 2.679 & 18.30 &  0.68 \\
03a60905     & 60904.8024 & $o$ &  4 & 18.84 & 16.80 & 16.26 & 16.03 & 0.01 & 2.953 & 2.642 & 19.86 &  0.72 \\
05r60908     & 60907.8897 & $w$ &  4 & 18.96 & 16.97 & 16.47 & 16.28 & 0.01 & 2.856 & 2.623 & 20.73 &  0.74 \\
04a60914     & 60914.0567 & $o$ &  4 & 18.44 & 16.40 & 15.88 & 15.65 & 0.01 & 2.666 & 2.593 & 22.11 &  0.78 \\
02a60916     & 60916.2574 & $o$ &  4 & 18.36 & 16.34 & 15.83 & 15.59 & 0.01 & 2.599 & 2.585 & 22.48 &  0.79 \\
03a60917     & 60916.7302 & $o$ &  4 & 18.41 & 16.37 & 15.82 & 15.59 & 0.01 & 2.584 & 2.583 & 22.55 &  0.79 \\
\label{tab:phot}
\end{longtable}

The magnitudes from Teide ($w$ band) are subject to revision, but we list the $w$ band results in Table~\ref{tab:phot} to provide knowledge of when ATLAS $w$ band exposures exist. Teide uses a complex system of 16 Celestron RASA-11 telescopes co-aligned in groups of 4, and it takes 10~sec sub-exposures.  The 12 sub-exposures for each field are all aligned and stacked into a single image, but the telescopes do not overlap perfectly, and the PSF can be quite complex because of wind buffeting and focus.  In particular, 3I/ATLAS was closer than 4\arcmin\ to the edge of the images on 05r60868 and 05r60883 and lack of coverage by all of the 16 sub-exposures may cause the photometric calibration to be in error.  The reduction pipeline for Teide is also still evolving, so if it proves to be important it is possible to re-reduce the Teide images of 3I/ATLAS to improve the accuracy of the $w$ band photometry, perhaps by using a subset of sub-exposures.

A way to judge the significance of the early detections is to ask how frequently a 6$\times$6\arcsec\ aperture in the stacked image is brighter than the one that is selected by the known orbit of 3I/ATLAS.  For the first stack epoch, encompassing MJD 60760--60766, the orbit selected pixel is exceeded 1.2\% of the time for both filters, 2.2\% of the time for the $c$ filter, and 4.4\% of the time for the $o$ filter.  Gaussian statistics would call these ``$2\sigma$ detections''.  For the second epoch, MJD 60786--60798, the orbit selected pixel is exceeded 0.04\% of the time for both filters combined, 1.3\% of the time for the $c$ filter, and 0.1\% of the time for the $o$ filter.  Gaussian statistics would call these ``$3\sigma$ detections''.  In fact the significance is better than this because we could condition consideration of false alarms by whether there is a bright blob in the stack before differencing and disregard those pixels (none of the early detection stacks has a bright blob at the location from the orbit).  The uncertainties in Table~\ref{tab:phot} are therefore fairly reliable indicators of false alarm probability as well as measurement uncertainty.  After the first two epochs the orbit centered pixels are much brighter than any other among $10^4$ trials and plainly visible to the eye.

\begin{figure}[h]
\begin{center}$
\begin{array}{cc}
\includegraphics[width=0.45\linewidth]{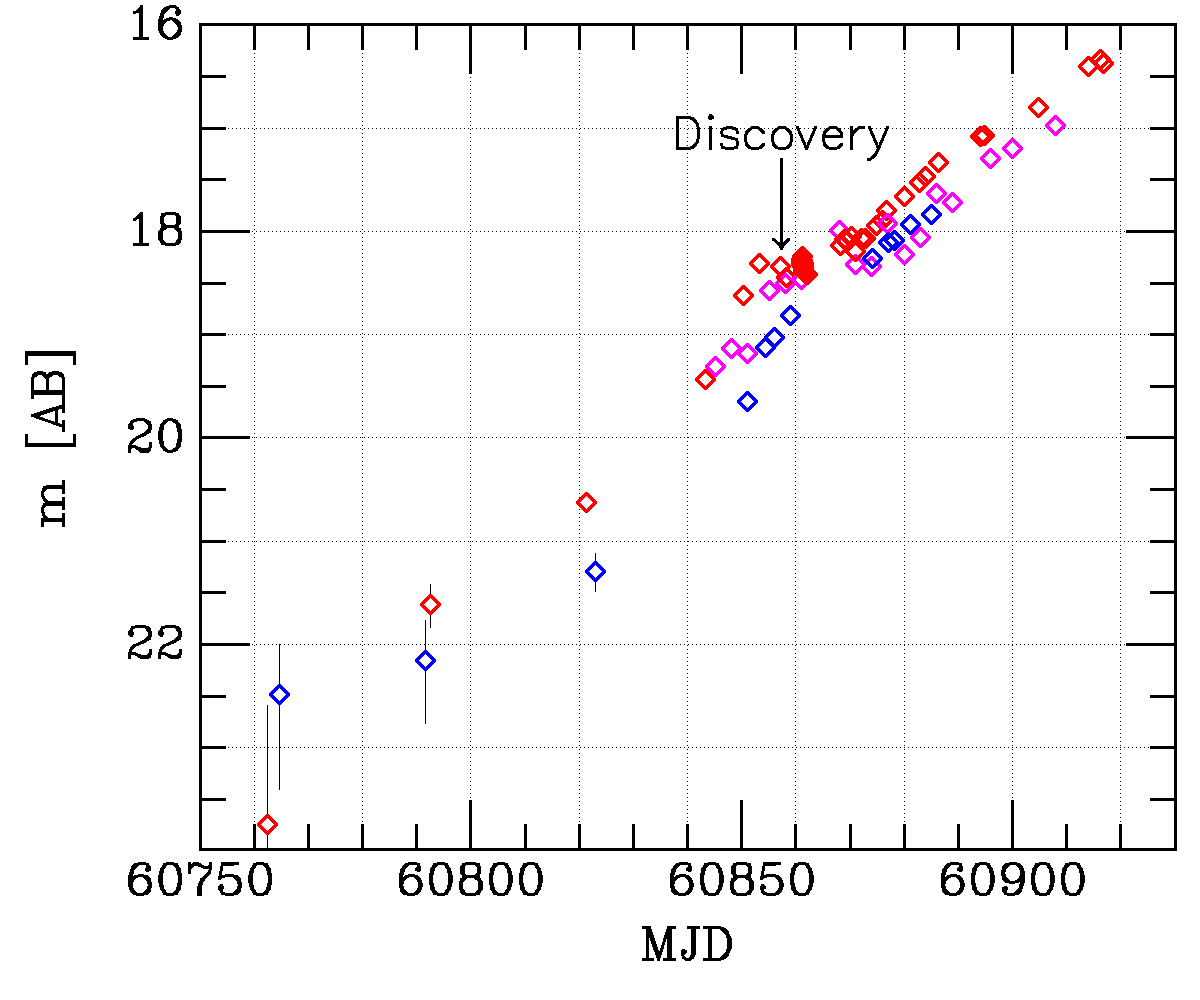}
\qquad
\includegraphics[width=0.45\linewidth]{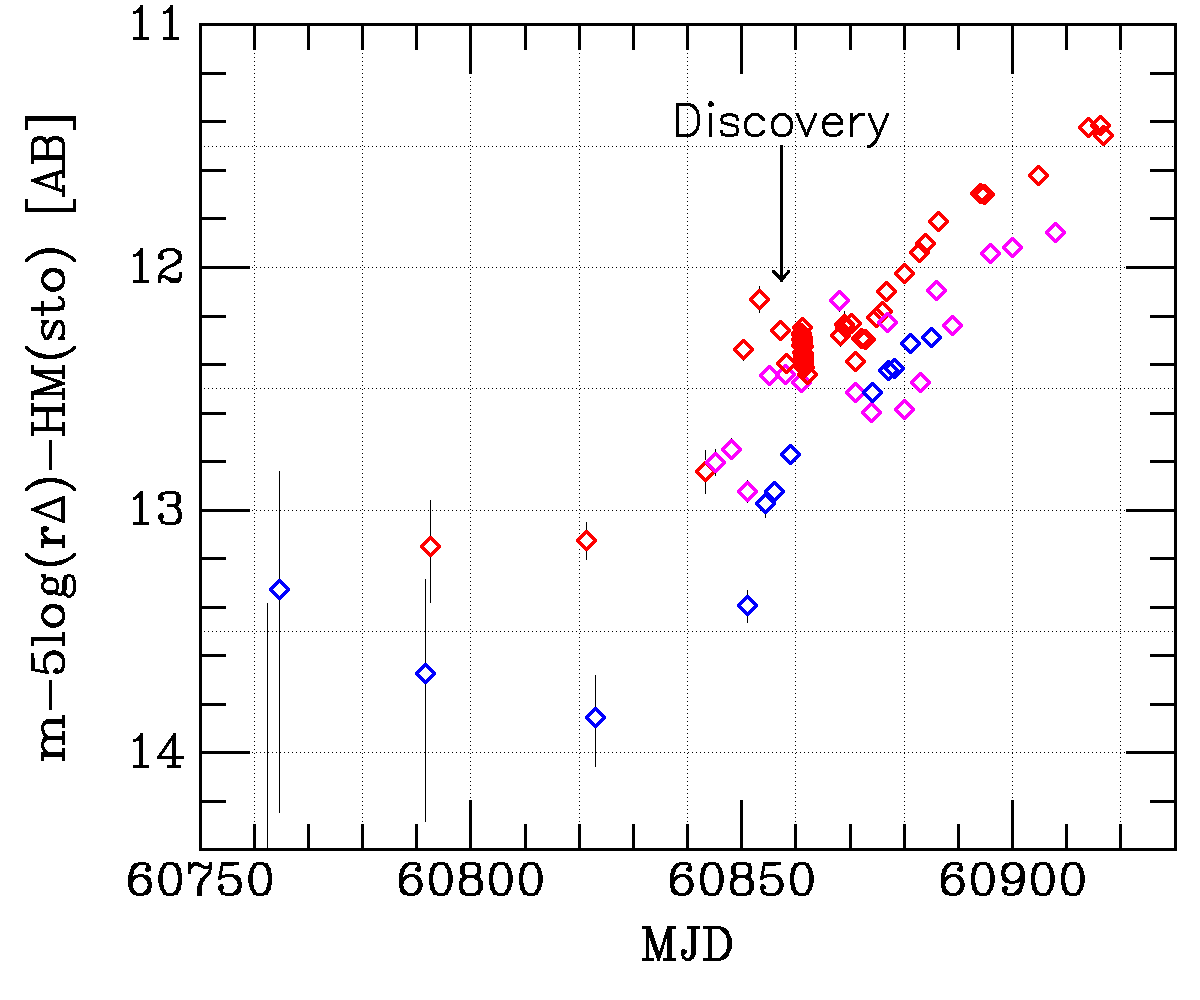}
\end{array}$
\end{center}
\caption{Magnitudes of 3I/ATLAS in a 6$\times$6\arcsec\ aperture are shown on the left as a function of time.  Red is $o$ band, blue is $c$ band, magenta is $w$ band from Teide. On the right the magnitudes are corrected for geometrical illumination and distance,  $5\log(r\Delta)$, and the Schleicher HM phase function, i.e. an absolute $H$ magnitude in the observed bandpass that shows the physical evolution of 3I/ATLAS with time.}
\label{fig:lc}
\end{figure}

Figure~\ref{fig:lc} illustrates the evolution of the brightness of 3I/ATLAS in a pair of graphs.  Left are the magnitudes as a function of time, right are the magnitudes corrected by distance from Sun $r$, Earth $\Delta$, and angular phase function to absolute $H(1,1,0)$ magnitude: the brightness of a fully illuminated object if at 1~au distance from both Sun and Earth (not physically realizable).

Of course it should be understood that most of the light at all times was coming from the extended coma, not the nucleus of this comet, evident in the brightening in successive apertures.  Therefore application of the HM phase function for these evolving coma particles to construct the $H$ magnitude may be misleading.  As seen in the table, the HM phase function is very nearly 0.04~mag/deg for the observed $4^\circ<\hbox{sto}<23^\circ$, but this can be revised as desired for a given physical model.  Similarly the volume encompassed by the apertures is a 2 dimensional pencil beam through at 3 dimensional coma and the geometrical correction must be interpreted with this in mind.

\section{Discussion and Conclusions}

The evolution of the absolute $H$ brightness of 3I/ATLAS is visible in the right panel of Figure~\ref{fig:lc}, although the two earliest epochs may not be fully reliable, as described above.  The $H$ magnitudes at 60792 and 60828 are distinctly fainter than the later points, contrary to the ZTF magnitude reported by MPEC 2025-N51\footnote{\tt https://minorplanetcenter.net/mpec/K25/K25N51.html}, converted to $H_V$ by \cite{Seligman2025}.
Since neither uncertainty nor aperture were reported for the ZTF magnitude, this may not be a real discrepancy.

Broadly speaking, there are a number of features visible from the light curve which are worth mentioning.
\begin{itemize}
  \item{} The early light curve (MJD$<60860$) appears quite red, $(o-c)\approx0.7$, and fairly flat in $H$ magnitude for two months prior to MJD 60845.
  \item{} Between 60845$<$MJD$<$60860 the color remained red and the $H$ magnitude brightened significantly and then declined.  This is the time of the discovery and the CFHT and VLT observations described in \cite{Seligman2025} of a solar-pointing ``rubble fountain''.
  \item{} The continuous, 12 hour sequence of observation on 60861 shows a steady decline in brightness by +0.27~mag/day and RMS residual of 0.025~mag that is consistent with the uncertainties; 
 no significant rotational modulation is visible.
  \item{} Since 60870 the color has been $(c-o)\approx0.3$, close to the solar value of $(c-o)=0.286$.
  \item{} The slope of $H(t)$ changed markedly around 60890 at heliocentric distance $\sim$3.3~au, from $-0.035$~mag/day to $-0.012$~mag/day, just about the time that 3I/ATLAS developed the anti-solar tail visible in Figure~\ref{fig:evolve}.  Expressed in terms of $r$, this corresponds to 
  a coma cross section that goes as $r^{-2}$ overall, but $r^{-3.9}$ prior to 60890 and $r^{-1.1}$ ever since.
\end{itemize}


ATLAS is hardly the only facility observing 3I/ATLAS, but we will not attempt to provide a summary of the observations to date (published or otherwise) because new papers appear on the arXiv every day and many fine datasets are still being carefully reduced and analyzed.  A thoughtful comparison and interpretation of all datasets is beyond the scope of these simple data, and since we now have the ATLAS results to date in a final state we think it appropriate to make them available to all as quickly as possible.

\begin{acknowledgments}
This work has made use of data from the Asteroid Terrestrial-impact Last Alert System (ATLAS) project. ATLAS is primarily funded to search for near-Earth asteroids through NASA grants NN12AR55G, 80NSSC18K0284, 80NSSC18K1575, and 80NSSC18K0265; byproducts of the NEO search include images and catalogs from the survey area.  The ATLAS science products have been  made possible through the contributions of the University of Hawaii Institute for Astronomy, the Queen's University Belfast, the Space Telescope Science Institute, the South African Astronomical Observatory (SAAO), and the Millennium Institute of Astrophysics (MAS), Chile.
ATLAS-Teide is an IAC instrument included in the present “Strategic plan of the Canarian Observatories”, funded by the European Union - NextGenerationEU EQC2021-007122-P project and the Spanish "Ministerio de Ciencia e Innovaci\'on" action ICT2022-007828.  AC acknowledges the support from ANID, Chile, through grants AIM23-0001 (Millennium Science Initiative) and FONDECYT No. 1251692.

\end{acknowledgments}

\bibliography{3I}{}
\bibliographystyle{aasjournal}

\end{document}